\documentclass[a4paper,fleqn,usenatbib]{mnras}

\usepackage{newtxtext,newtxmath}

\usepackage{newtxtext,newtxmath}
\usepackage{graphicx}	
\usepackage{amsmath}	
\usepackage{amssymb}	
\usepackage{newtxtext,newtxmath}
\usepackage[T1]{fontenc}
\usepackage{ae,aecompl}
\usepackage{subfigure}
\usepackage{subfloat}
\usepackage{multirow}
\usepackage{multicol}
\usepackage{lipsum}
\usepackage{dcolumn}
\usepackage{bm}

\title{Principle of Helical \& Nonhelical Dynamo and $\alpha$ effect in Field Structure model}

\author[Kiwan Park]{
Kiwan Park,$^{1}$\thanks{E-mail: oz150@uni-heidelberg.de, pkiwan@gmail.com}
\\
$^{1}$Center for Astronomy, Institute for Theoretical Astrophysics at University of Heidelberg 69120 Heidelberg, Germany}

\date{Accepted XXX. Received YYY; in original form ZZZ}

\pubyear{2018}

\begin{document}
\label{firstpage}
\pagerange{\pageref{firstpage}--\pageref{lastpage}}
\maketitle


\begin{abstract}
We explain the (non)helical dynamo process using a field-structure model based on magnetic induction equation in an intuitive way. We show how nonhelical kinetic energy converts into magnetic energy and cascades toward smaller eddies in a mechanically forced plasma system. Also, we show how helical magnetic energy is inversely cascaded ($\alpha$ effect) toward large scale magnetic eddies in a mechanically or magnetically forced system. We, then, compare the simulation results with the model qualitatively for the verification of the model. In addition to these intuitive and numerical approaches, we show how to get $\alpha$ and $\beta$ coefficient semi-analytically from the temporally evolving large scale magnetic energy and magnetic helicity.
\end{abstract}





\section{Introduction}
Although the various scales of magnetic field ${\mathbf{B}}$ and conducting fluids (plasma) are ubiquitous in space, it is not yet clearly understood how the magnetic fields and plasmas exchange energy through their mutual interactions \citep{1978mfge.book.....M, 1980opp..bookR....K, 2005PhR...417....1B}. The energy transferred from the conducting fluid to the magnetic field generates various scales of magnetic fields and amplifies them (dynamo). Briefly, the dynamo phenomena are classified as a large-scale dynamo (LSD) and a small-scale dynamo (SSD) depending on the direction of energy transfer. In particular, since many physical turbulent phenomena e.g. transport of momentum or material are mostly controlled by large scale motions, the evolution and role of large scale magnetic field ($\overline{B}$) in a turbulent plasma system are fundamental and practical problems that cannot be limited only to academic interests.\\

Large scale dynamo theory shows how the small scale magnetic energy with helicity ($\alpha$ effect, \citep{2012MNRAS.419..913P, 2012MNRAS.423.2120P}), differential rotation ($\Omega$ effect, \cite{1991ApJ...376..214B}), or shear current (\cite{2003PhRvE..68c6301R}) can be (inversely) cascaded toward ${\overline{\mathbf{B}}}$. Out of them, the $\alpha$ effect is indispensable to the self consistent dynamo process, or inverse cascade of magnetic energy $E_M$ in the helical large scale dynamo. Moreover, since the properties of the helicity provide a relatively clear mathematical advantage in the theoretical description of the LSD phenomena, many LSD theories aim to represent electromotive force (EMF, $\xi \equiv \langle \mathbf{u}\times \mathbf{b}\rangle$), which is a source of $\overline{\mathbf{B}}$, with (pseudo) tensors and $\overline{\mathbf{B}}$.\\

Analytically $\alpha$ effect can be derived with a scale-divided function feedback method which is also a basic principle of numerical calculation. The representative theories like first order smoothing approximation (FOSA, or second order correlation approximation, SOCA, \cite{1978mfge.book.....M, 1980opp..bookR....K, 2005PhR...417....1B}), minimal tau approximation (MTA, \cite{2002PhRvL..89z5007B}), or Quasi Normalized approximation (QN, \cite{1975JFM....68..769F}) are actually based on the method in a dynamic or stationary state. However, since the $\alpha$ effect is not a strict mathematical concept, some ambiguities are inevitable for the analytical derivation, which does not depreciate the importance of $\alpha$ effect in the helical dynamo. Numerically, $\alpha$, $\beta$ coefficient in the $\alpha$ effect can be found applying an external magnetic field $\mathbf{B}_{ex}$ to the system (\cite{2005AN....326..245S}). However, since $\mathbf{B}_{ex}$ affects the dynamo process (\cite{1996PhRvE..54.4532C}), well-designed numerical results without $\mathbf{B}_{ex}$ would be used to derive the coefficients (test field method, \cite{2005AN....326..245S}).\\

Conventional dynamo theories show what happen over the statistical number of realizations of magnetohydrodynamic (MHD) system. The analytical theories give a qualitative and more or less quantitative description of the evolution of magnetic fields in the plasma, but they do not tell us how the actual plasma and magnetic field interact physically within the system. To explain the physical processes of evolving $\mathbf{B}$, a cartoon model of stretching of $\mathbf{B}$ field `($\mathbf{B}\cdot \nabla) \mathbf{u}$' has been used in analogy to stretching vorticity ($\mathbf{\omega}\cdot \nabla \mathbf{u}$, $\mathbf{\omega}=\nabla\times \mathbf{u}$) neglecting tilting effect \citep{1983mfa..book.....Z, 2002ApJ...576..806S}. However, $\mathbf{B}$ is essentially not so directly related to $\mathbf{u}$ as $\mathbf{\omega}$ is. Moreover, the concept of `stretching, twist, folding' is not relevant to any physics law or fluidal equation. Furthermore, the model implying the co-stretching of $\mathbf{B}$ and $\mathbf{u}$ needs to explain the nontrivial EMF ($\sim\langle \mathbf{u}\times \mathbf{B}\rangle \neq 0$). This large gap between the dynamo model and the dynamo mechanism makes it more difficult to derive a more accurate dynamo theory.
\\

Here, we introduce an improved field structure model \citep{2017MNRAS.472.1628P} based on magnetic induction equation for the physical mechanisms of a helical large scale dynamo (LSD) and nonhelical small scale dynamo (SSD). The dynamo processes shown in the model are in line with the theory and consistent with the simulation results. Also, we show how to get $\alpha$ coefficients in the helical LSD from large scale magnetic energy $\overline{E}_M\,(\langle \overline{B}^2\rangle /2)$ and magnetic helicity $\overline{H}_M\,(\langle \overline{\mathbf{A}}\cdot \overline{\mathbf{B}}\rangle )$ which can be measured in observation and simulation.

\section{Simulation}
For the numerical investigation we used the $\mathrm{PENCIL}$ $\mathrm{CODE}$. We used $\mathrm{PENCIL}$ $\mathrm{CODE}$, which solves the coupled fluid equations for the compressible conducting fluids in a periodic box \citep{2001ApJ...550..824B}.
\begin{eqnarray}
\frac{D \rho}{Dt}&=&-\rho {\bf \nabla} \cdot {\bf U},\label{continuity equation for pencil code}\\
\frac{D {\bf U}}{Dt}&=&-{\bf \nabla} \mathrm{ln}\, \rho + \frac{1}{\rho}(\nabla\times{\bf B})\times {\bf B}\nonumber\\
&&+\nu\big({\bf \nabla}^2 {\bf U}+\frac{1}{3}{\bf \nabla} {\bf \nabla} \cdot {\bf U}\big)\,\, (+\mathbf{f}_{kin})
\label{momentum equation for pencil code}\\
\frac{\partial {\bf A}}{\partial t}&=&{\bf U}{\bf \times} {\bf B} -\eta\,{\bf \nabla}{\bf \times}{\bf B}\,\,(+\mathbf{f}_{mag}).\label{magnetic induction equation for pencil code}
\end{eqnarray}
Here, $\rho$ and $D/Dt(=\partial / \partial t + {\bf U} \cdot {\bf \nabla}$) indicate the density and Lagrangian time derivative. $\nu$ and $\eta$ are kinematic viscosity and magnetic diffusivity respectively. The velocity is in unit of the sound speed, and the magnetic field is normalized by $(\rho_0\,\mu_0)^{1/2}c_s$, where $\mu_0$ and $c_s$ are magnetic permeability and sound speed, respectively. The forcing function ${\bf f}(x,t)$ in Fourier space is $N\,{\bf f}_k(t)\, exp\,[i\,{\bf k}_f(t)\cdot {\bf x}+i\phi(t)]$:
\begin{eqnarray}
{\bf f}_{kin,\,or mag}(t)=\frac{i\mathbf{k}(t)\times (\mathbf{k}(t)\times \mathbf{e})-\xi |k(t)|(\mathbf{k}(t)\times \mathbf{e})}{k(t)^2\sqrt{1+\xi^2}\sqrt{1-(\mathbf{k}(t)\cdot \mathbf{e})^2/k(t)^2}}.\nonumber
\label{forcing ampliitude fk}
\end{eqnarray}
Here `$\mathbf{e}$' is an arbitrary unit vector, `$\xi$' denotes the helicity ratio, and `$\phi(t)$' is a random phase ($|\phi(t)|\leq\pi$). For example if `$\xi$' is `$\pm1$', $i\mathbf{k}\times \mathbf{f}_k=\pm k\mathbf{f}_k$ (fully helical). If `$\xi$' is `0', $i\mathbf{k}\times \mathbf{f}_k$ is not proportional to $\mathbf{f}_k$. The calculated forcing function in Fourier space at $k=k_f$ is again inversely Fourier transformed and applied to Eq.~(\ref{momentum equation for pencil code}) or (\ref{magnetic induction equation for pencil code}).\\

We simulated MHD systems forced with a helical kinetic forcing dynamo (HKFD), helical magnetic forcing dynamo (HMFD), and nonhelical kinetic forcing dynamo (NHKFD) with an isothermal environment ($\pi^3$, a periodic boundary condition). For the HKFD, a fully helical energy ($\nabla\times\mathbf{f}_{kin}=k_f\mathbf{f}_{kin}$, $k_f=5$) is given to  Eq.~(\ref{momentum equation for pencil code}). For the HKFD, a fully helical energy ($\nabla\times\mathbf{f}_{mag}=k_f\mathbf{f}_{mag}$) is given to Eq.~(\ref{magnetic induction equation for pencil code}), but $|\mathbf{f}_{mag}|=|\mathbf{f}_{kin}|/k_f$. For NHKFD, the forcing method is the same as that of HKFD except the helicity ratio: $\nabla\times\mathbf{f}\nsim\mathbf{f}$.\\

\begin{figure}
     \includegraphics[width=7.5cm]{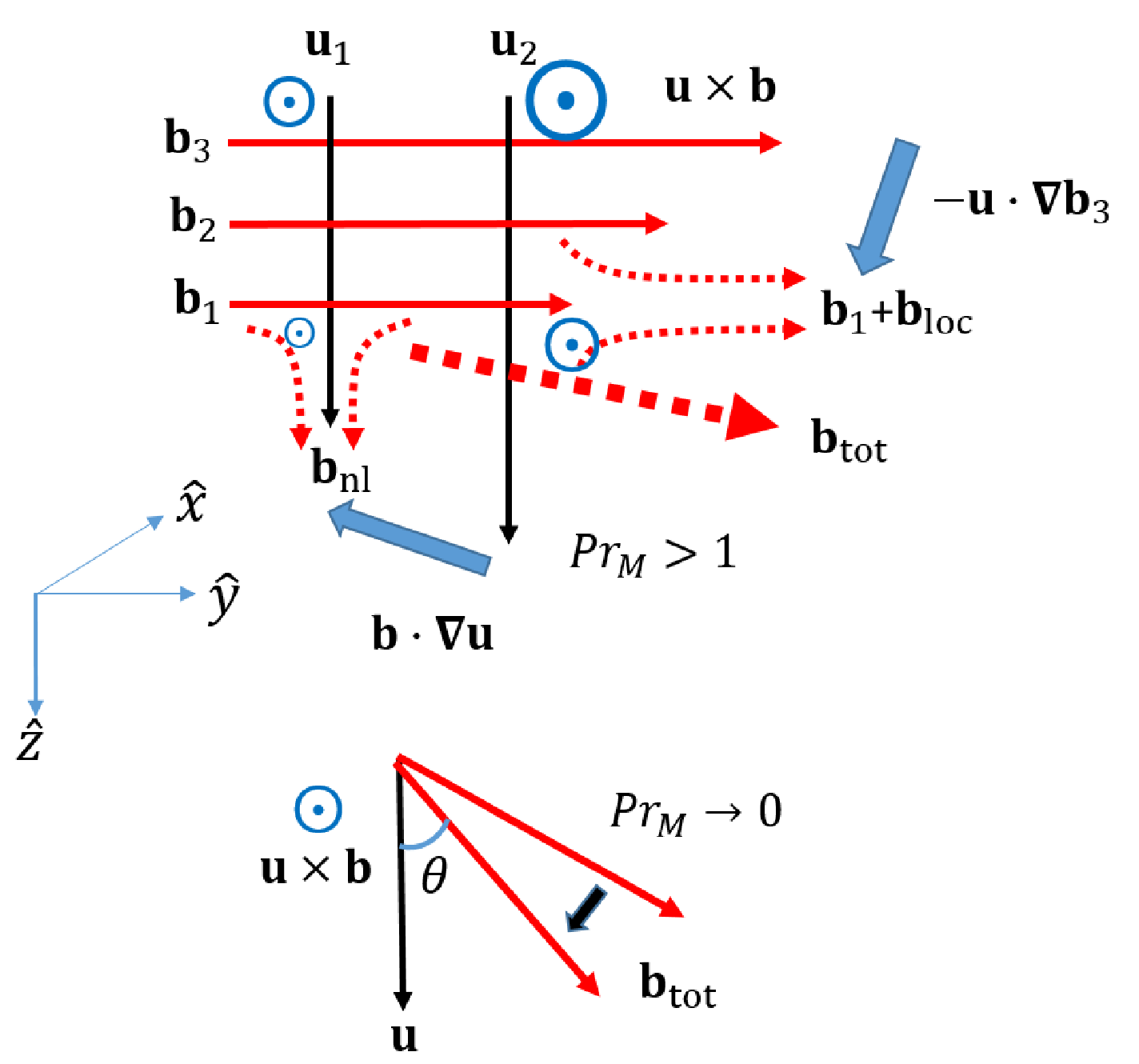}
\label{f1}
\caption{Field structure from $\mathbf{b}_i\cdot \nabla \mathbf{u}_i-\mathbf{u}_i\cdot \nabla \mathbf{b}_i\,(\sim\nabla \times \langle \mathbf{u}_i\times \mathbf{b}_i\rangle)$ shows how nonhelical magnetic energy migrates along $\mathbf{u}_i$ and $\mathbf{b}_i$.} 
\end{figure}

\begin{figure*}
\centering{
  {
   \subfigure[]{
     \includegraphics[width=6.2cm]{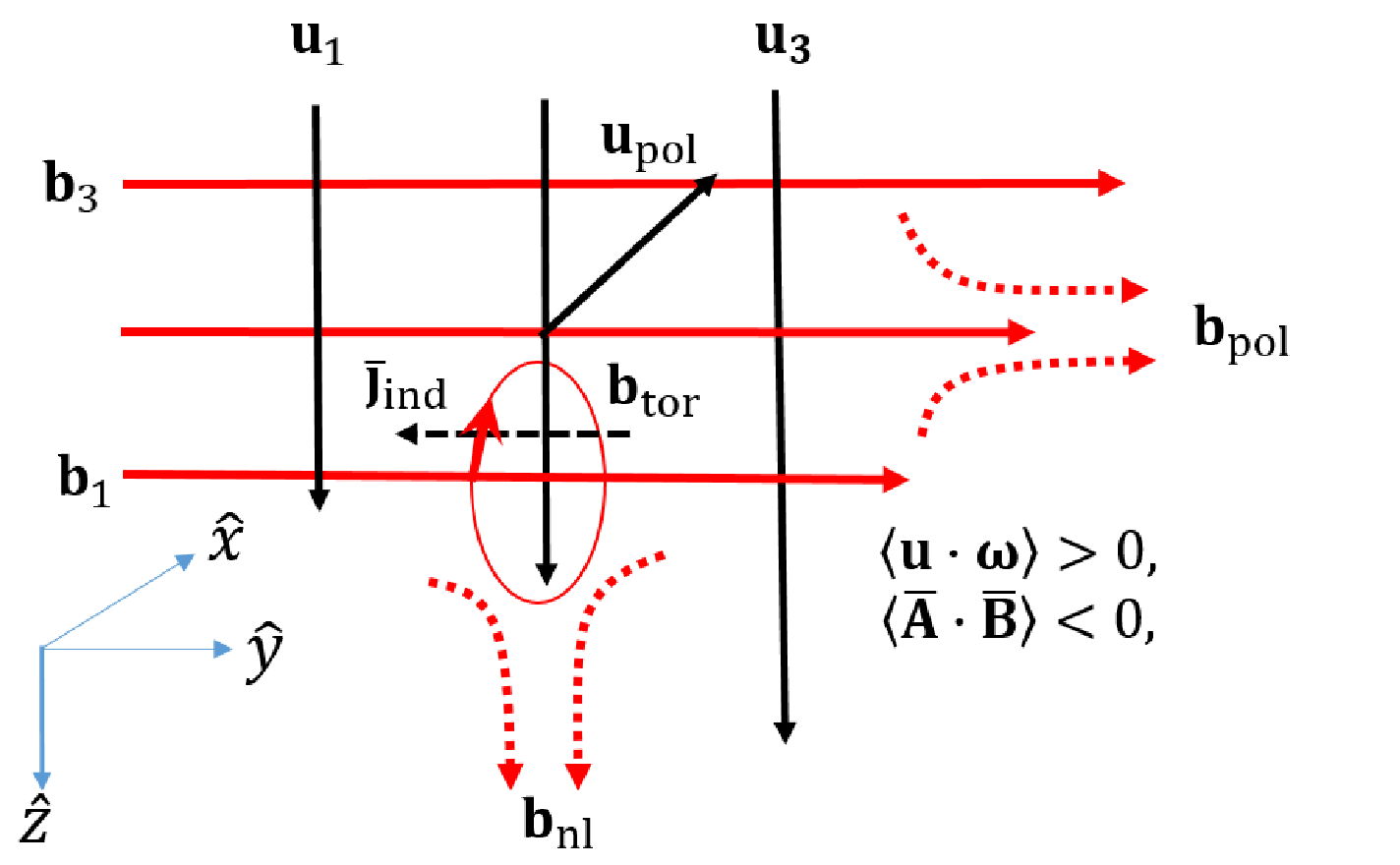}
     \label{f2}
   }\hspace{0mm}
   \subfigure[]{
     \includegraphics[width=6.9cm]{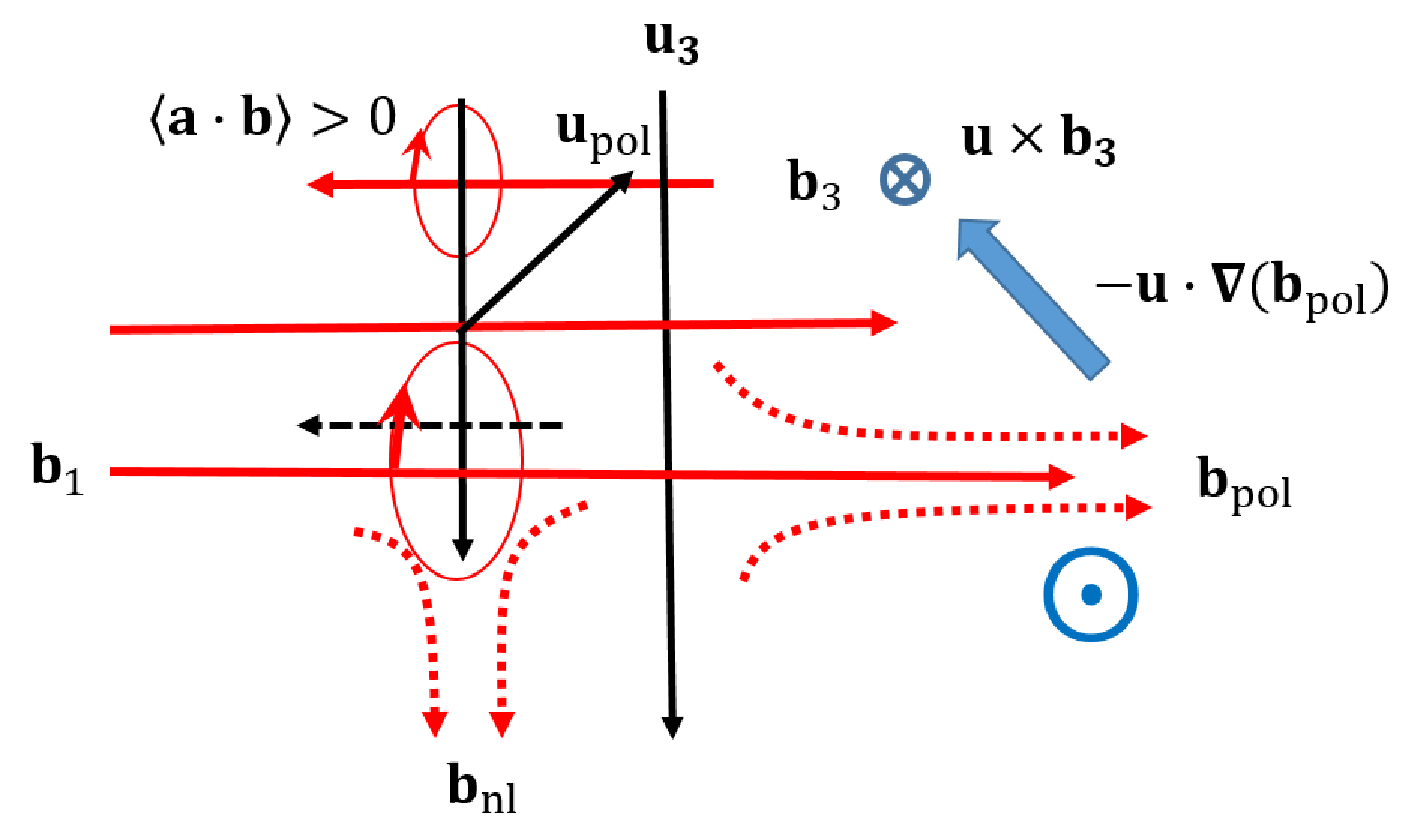}
     \label{f3}
   }
   }
\caption{(a) $\mathbf{b}_{tor}$ and $\mathbf{b}_{pol}$ form a left handed (negative) magnetic helicity, and $\mathbf{u}_{tor}$ and $\mathbf{u}_{pol}$ form a right handed (positive) kinetic helicity. (b) Magnetic energy at $\mathbf{b}_{1}$ diffuses toward $\mathbf{b}_{3}$ through $-\mathbf{u}\cdot\nabla \mathbf{b}$.}
}
\end{figure*}

\begin{figure*}
\centering{
  {
   \subfigure[\,HKF]{
     \includegraphics[width=8cm]{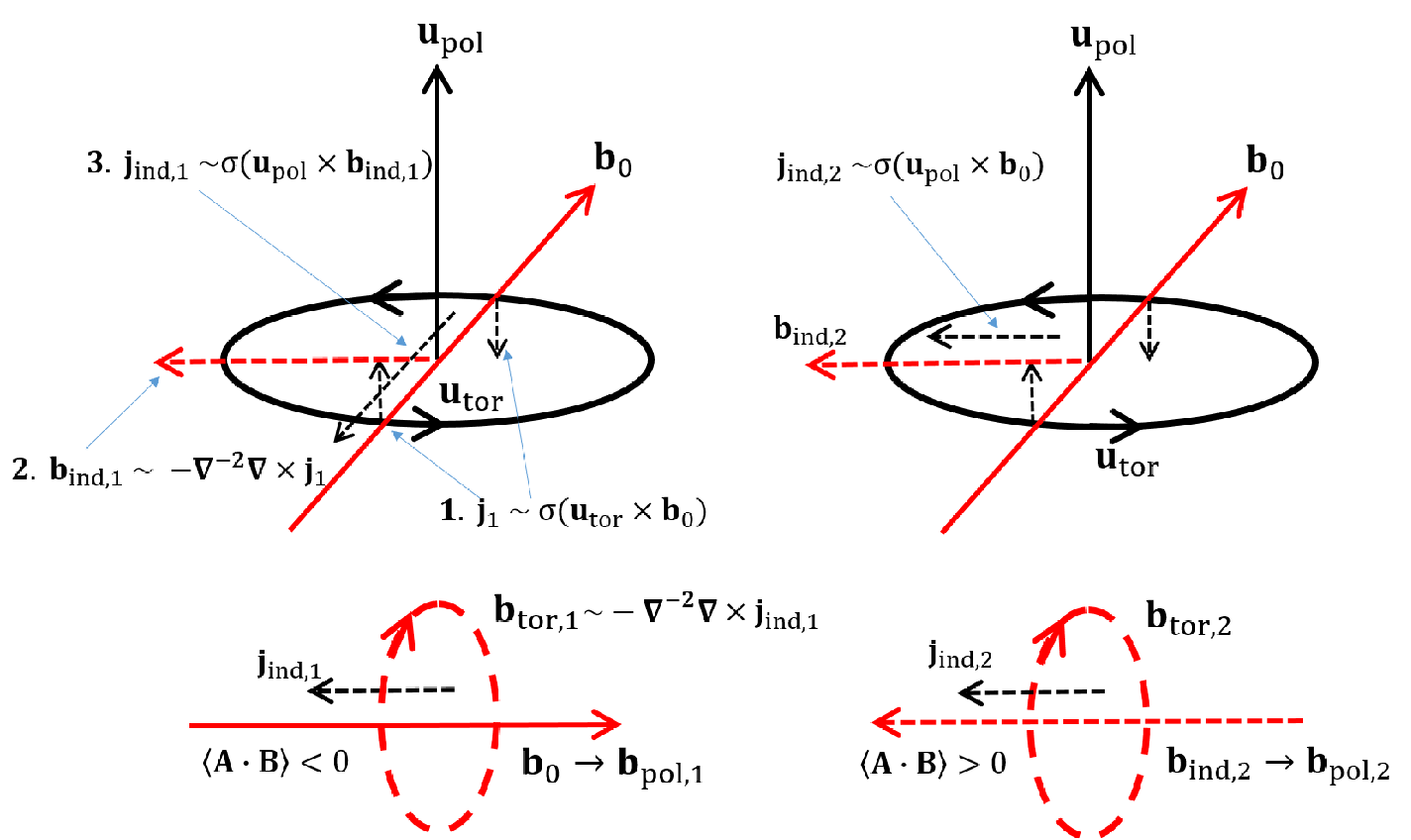}
     \label{f4a}
   }\hspace{5mm}
   \subfigure[\,HMF]{
     \includegraphics[width=8.1cm]{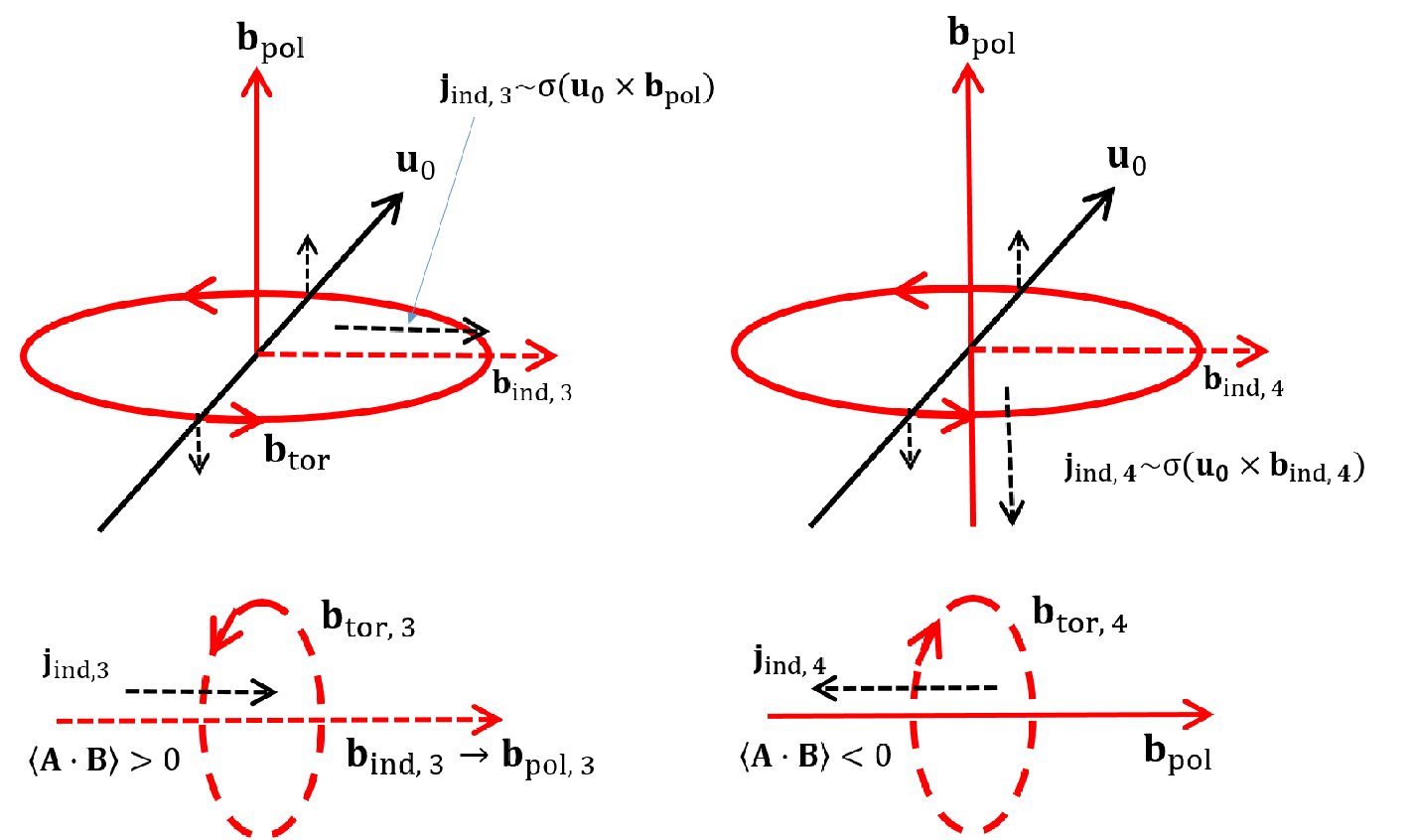}
     \label{f4b}
   }
   }
\caption{In principle, a helically forced system (HKF or HMF) can generate both a positive and negative helical magnetic helicity.}
}
\end{figure*}

\begin{figure*}
\centering{
  {
   \subfigure[]{
     \includegraphics[width=6.15cm]{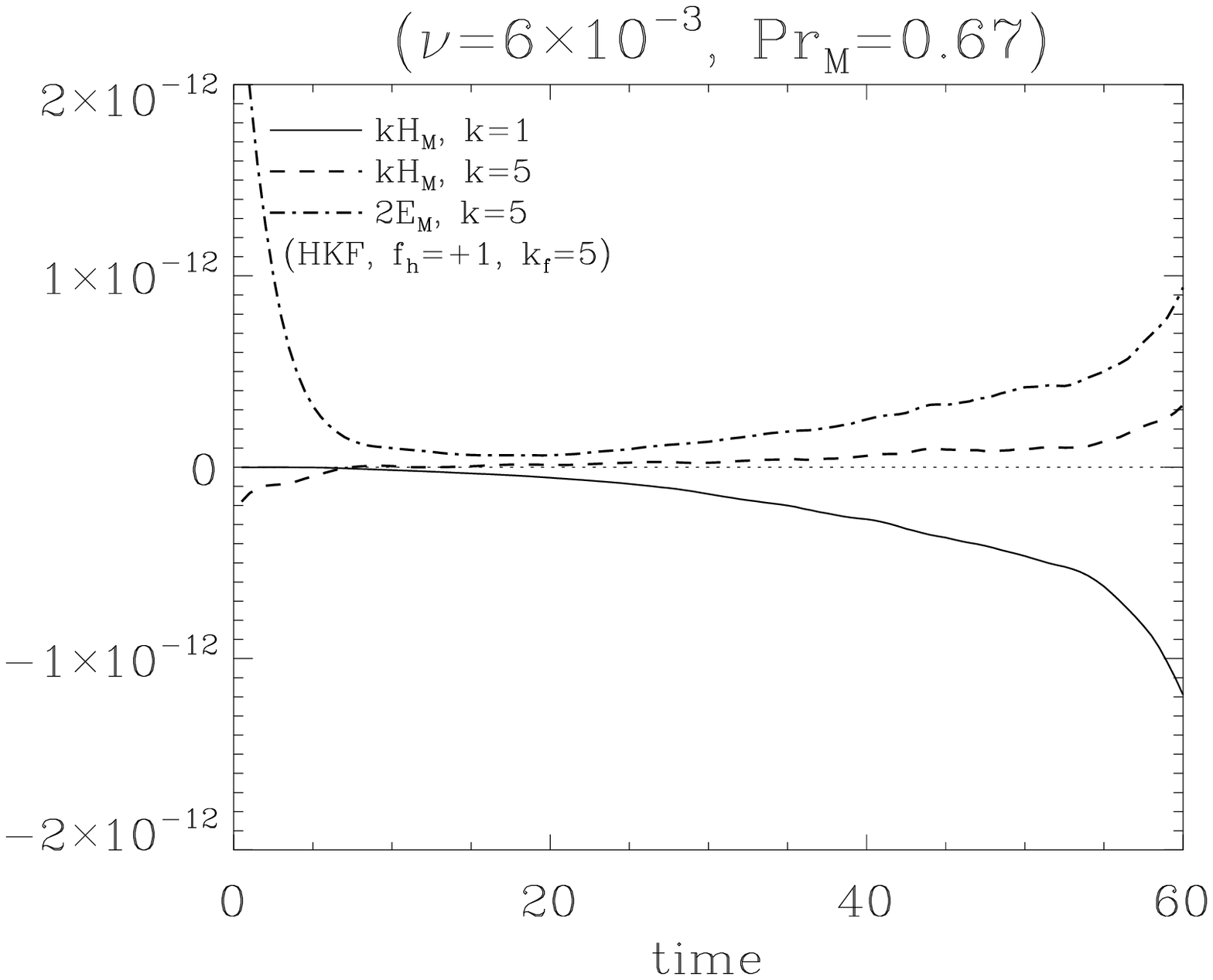}
     \label{f5a}
   }\hspace{-8mm}
   \subfigure[]{
     \includegraphics[width=6.15cm]{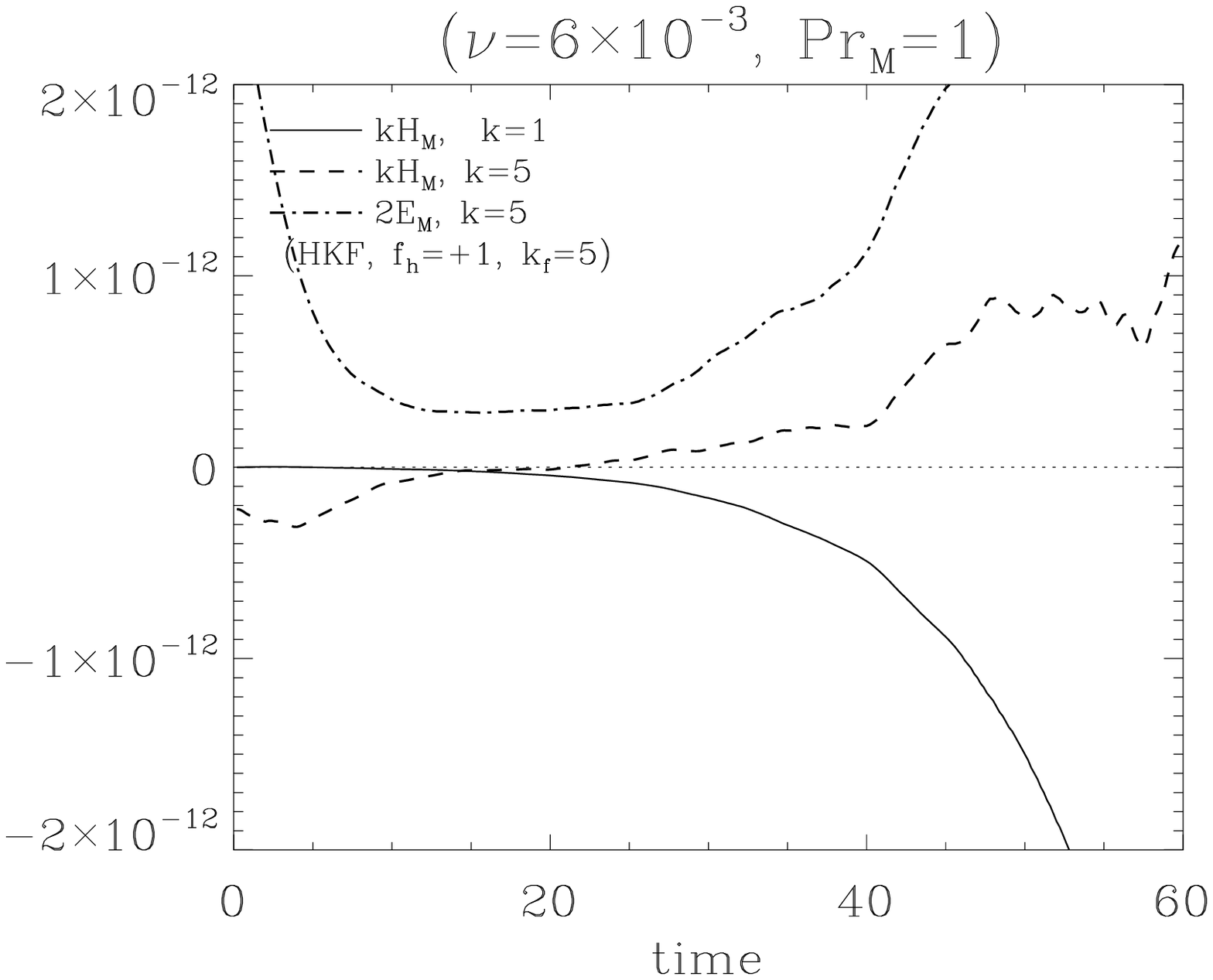}
     \label{f5b}
   }\hspace{-8mm}
   \subfigure[]{
     \includegraphics[width=6.2cm]{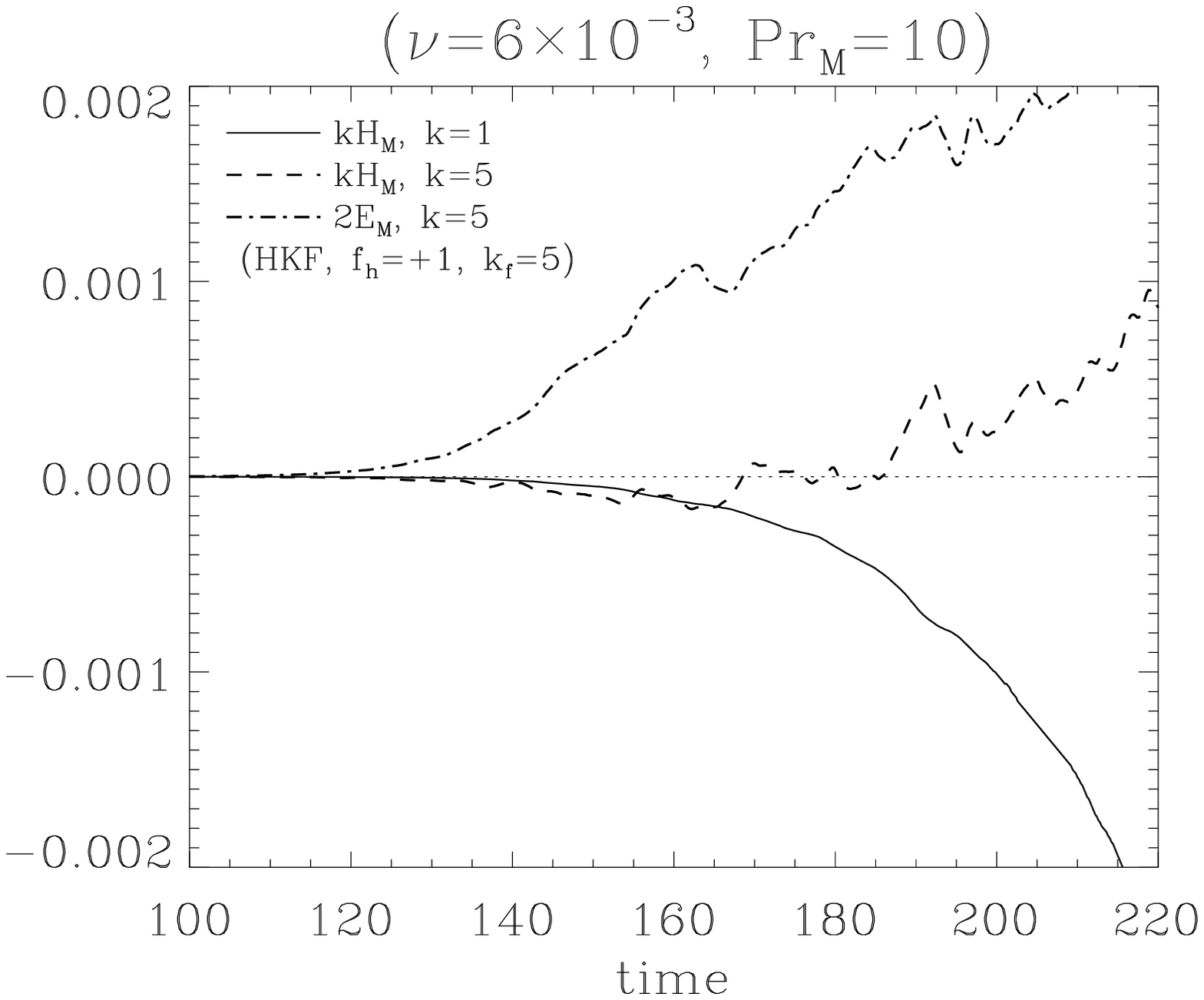}
     \label{f5c}
   }
   }
\caption{Evolving $E_M$ and $H_M$ at k=1, 5 in a system forced with the positive kinetic helicity. As $Pr_M$ increases, the forced eddy $k_f$ is relatively more influenced by the kinetic helicity than the magnetic diffusion from the large scale magnetic field.}
}
\end{figure*}

\begin{figure*}
\centering{
  {
   \subfigure[\,\,HKF \& HMF of $f_h=+1$]{
     \includegraphics[width=7.5 cm]{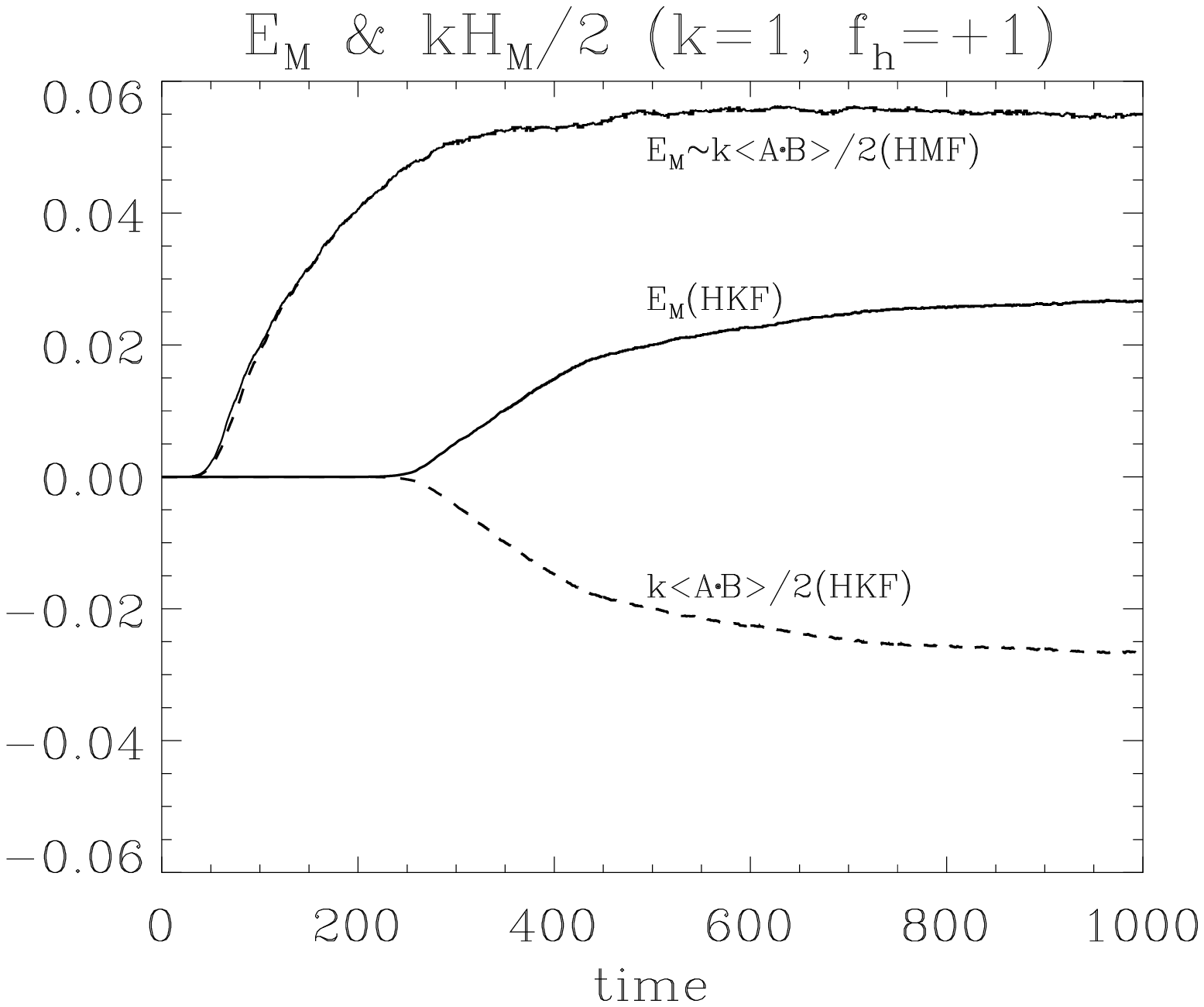}
     \label{f6a}
   }\hspace{2 mm}
   \subfigure[\,\,HKF \& HMF of $f_h=-1$]{
     \includegraphics[width=7.5cm]{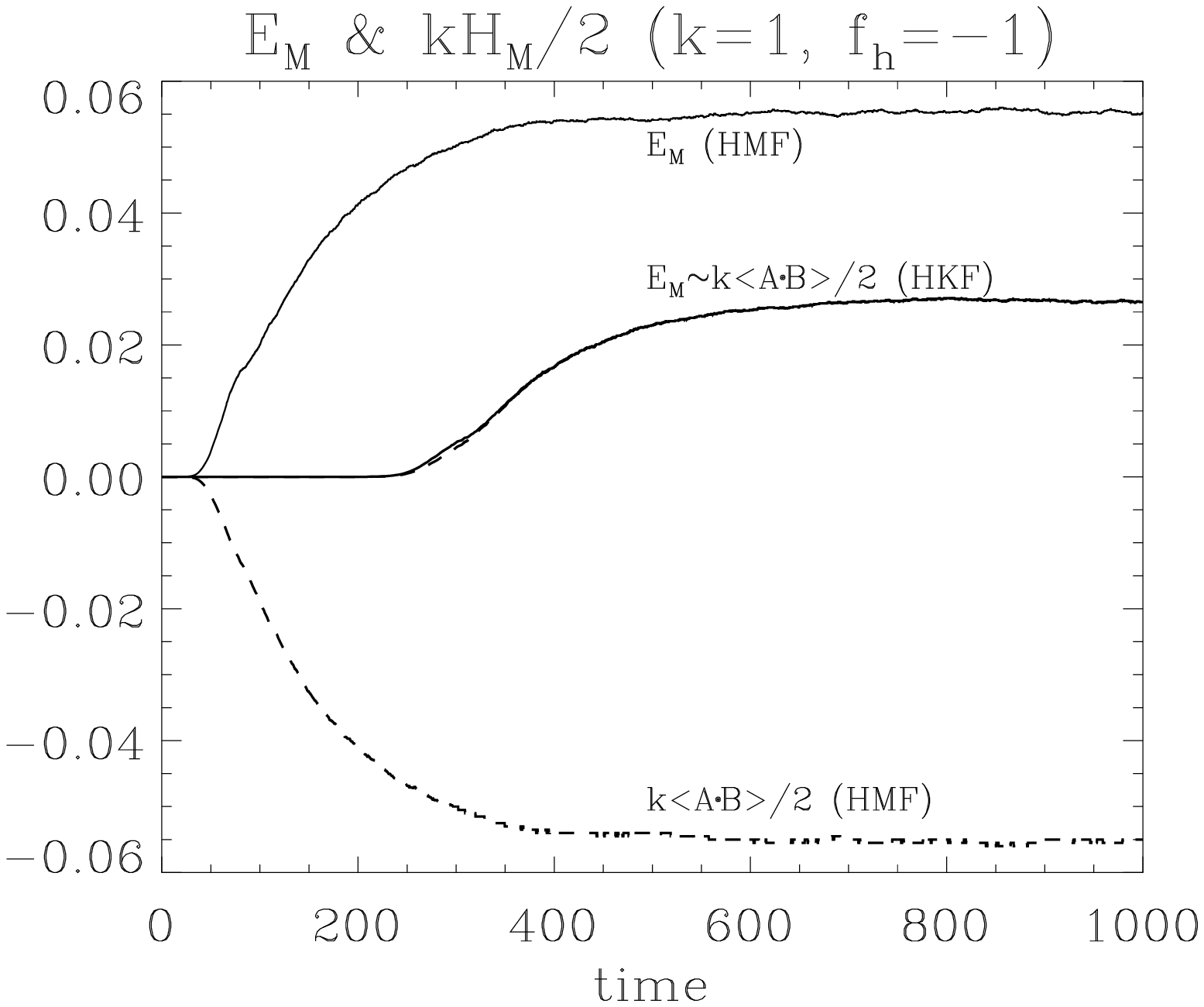}
     \label{f6b}
   }
   }
\caption{ (a), (b) Large scale magnetic energy $\overline{E}_M (\langle \overline{B}^2\rangle/2)$ and its helical component $k\langle\overline{\mathbf{A}}\cdot \overline{\mathbf{B}}\rangle/2\,(k=1)$ for HKFD and HMFD. The evolution of $E_M$ and $H_M$ can be explained by Eq.~(\ref{Hm_final_solution}), (\ref{Em_final_solution}) according to $\alpha>0$ or $\alpha<0$.}
}
\end{figure*}

\begin{figure}
\centering{
   \subfigure[\,\,NHKF ($f_h=0$)]{
     \includegraphics[width=9 cm]{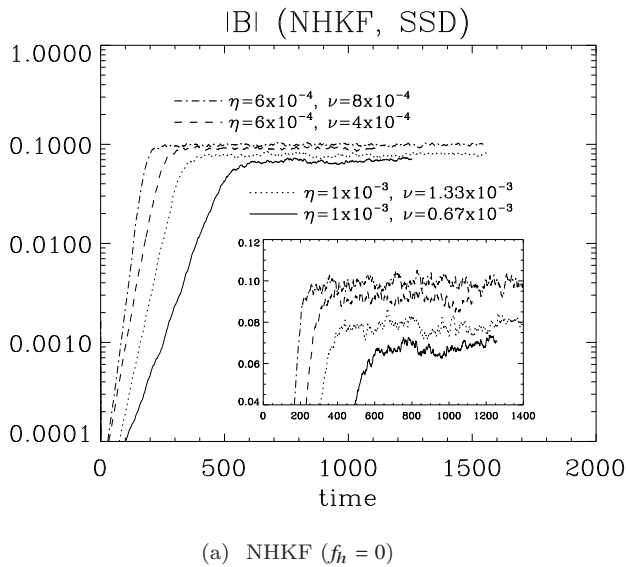}
     \label{f6c}
   }
\caption{ SSD forced with the nonhelical kinetic energy at $k$=5. $|B|$ is (inversely) proportional to $(\eta)\, \nu$ or magnetic Prandtl number $Pr_M$.}
}
\end{figure}


%
%
%

Dynamo process shown in Fig.~1 is mathematically straightforward. The interaction between $\mathbf{u}_i$ and $\mathbf{b}_i$ yields EMF ($\xi_i\sim\langle \mathbf{u}_i\times \mathbf{b}_i\rangle$($-\hat{x}$)), which is the weakest at $\mathbf{u}_1$ \& $\mathbf{b}_1$ and strongest at $\mathbf{u}_2$ \& $\mathbf{b}_3$. This inhomogeneous EMF arouses a nontrivial curl effect, i.e., growth rate of magnetic field: $\nabla \times \langle \mathbf{u}_i \times \mathbf{b}_i\rangle\sim \partial \mathbf{b}_i/\partial \, t$. The growth rate is the weakest at $\mathbf{u}_2$ \& $\mathbf{b}_3$ and strongest at $\mathbf{u}_1$ \& $\mathbf{b}_1$. This indicates energy transfer from $\mathbf{b}_3$ ($\mathbf{u}_2$) to $\mathbf{b}_1$ ($\mathbf{u}_1$). Essentially the transport of magnetic energy occurs in a bimodal way. A transferred magnetic field from $\mathbf{u}_2$ to $\mathbf{u}_1$ through $\mathbf{b}\cdot \nabla \mathbf{u}$\footnote{Here, a bold letter $\mathbf{u}$ or $\mathbf{b}$ indicates a vector field.} is represented by $\mathbf{b}_{nl}$ ($\mathbf{b}_{nl}(t)=\int^t \mathbf{b}\cdot \nabla \mathbf{u}\,d\tau$), which is parallel to the velocity field. Similarly, a transferred magnetic field from $\mathbf{b}_3$ to $\mathbf{b}_1$ through $-\mathbf{u}\cdot \nabla \mathbf{b}$ is represented by $\mathbf{b}_{loc}$ ($\mathbf{b}_{loc}(t)=-\int^t \mathbf{u}\cdot \nabla \mathbf{b}\,d\tau$), which is parallel to the magnetic field. Net magnetic field $\mathbf{b}_{tot}$ from these two transferred magnetic fields again interacts with $\mathbf{u}$ at the next dynamo process.\\


In addition to the strength of $\mathbf{u}$ and $\mathbf{b}$, the angle $\theta$ between $\mathbf{b}_{nl}$ and $\mathbf{b}_1+\mathbf{b}_{loc}$ plays a crucial role in EMF. If $\mathbf{b}_{nl}$ grow faster than $\mathbf{b}_{loc}$, $\theta$ and EMF decrease. Several factors can affect their relative ratio, but magnetic Prandtl number $Pr_M=\nu/\eta$ has a paradoxical effect. Both $\nu$ and $\eta$ are related to the dissipation of energy; but, their roles work in the opposite way. Decreasing $\eta$ increases $\mathbf{b}_{loc}$ and EMF. However, with small $\nu$ more kinetic energy is transported to a smaller eddy. Then, more magnetic energy is transferred to the smaller eddy leading to the growth of $\mathbf{b}_{nl}$, which disturbs dynamo (see Fig.~\ref{f6c}).\\

In the model $\mathbf{b}_1$ (or $\mathbf{u}_1$) can be considered as a large or small scale field. However, as $\partial \mathbf{b}/\partial t\sim \nabla \times \langle \mathbf{u} \times \mathbf{b}\rangle\sim l^{-1}\langle \mathbf{u} \times \mathbf{b}\rangle$ implies, if $\mathbf{u}_1$ and $\mathbf{b}_1$ are the small scale eddies, the magnetic energy transfer from $\mathbf{b}_3$ ($\mathbf{u}_3$) to $\mathbf{b}_1$ ($\mathbf{u}_1$) occurs more easily due to the small characteristic length. In contrast, if $\mathbf{b}_1$ and $\mathbf{u}_1$ are large scale eddies, additional process is required to overcome the large characteristic length $l_1$. Nonetheless, the energy transport is essentially bidirectional, which appears clearly in a decaying MHD system \citep{2017PhRvD..96h3505P, 2017MNRAS.472.1628P}.\\


\subsection{Amplification of the helical $B$ field for LSD}
Fig.~2 shows a dynamo system forced by the right handed helical kinetic energy (HKF) composed of a toroidal field $\mathbf{u}_{tor}$ and poloidal one $\mathbf{u}_{pol}(\hat{x})$ ($\langle \mathbf{u}\cdot \mathbf{\omega}\rangle>0,\, \omega=\nabla\times \mathbf{u})$. \footnote{In Fig.~\ref{f2} only a representative $\mathbf{u}_{pol}$ is considered to synchronize the two-scale dynamo theory. $\mathbf{j}_{ind}$, $\mathbf{b}_{tor}$, and the resultant left handed magnetic helicity exist in the whole scales.} The interaction between $\mathbf{u}_{pol}(\hat{x})$ and $\mathbf{b}_{nl}(\hat{z})$ induces a current density $\mathbf{j}_{ind}=\sigma(\mathbf{u}_{pol}\times \mathbf{b}_{nl})(-\hat{y})$, which generates a toroidal magnetic field $\mathbf{b}_{tor}(=-\nabla^{-2}\nabla\times \mathbf{j}_{ind})$ around $\mathbf{b}_1$(=$\mathbf{b}_{pol}$) forming a left handed magnetic helicity $(\langle\mathbf{a}\cdot \mathbf{b}\rangle<0)$. $\mathbf{b}_{tor}$ interacts with $\mathbf{u}$ to induce another circular current density $\mathbf{j}_{circ}$ antiparallel to $\mathbf{b}_{tor}$. $\mathbf{j}_{circ}$ amplifies $\mathbf{b}_{pol}$, which amplifies $\mathbf{b}_{tor}$ back through the $\alpha^2$ dynamo process \citep{2017PhRvD..96h3505P, 2017MNRAS.472.1628P}. This compound process makes possible $|\mathbf{b}_1|$ surpasses $|\mathbf{b}_2|$ and $|\mathbf{b}_3|$. As $|\mathbf{b}_i|$ grows, dissipation effect plays a decisive role. So if $\mathbf{b}_1$ is a small scale field, its dissipation $\sim k_1^2 b_1$ ($k_1\gg1$) becomes larger than that of other eddies. Therefore $\mathbf{b}_1$ should be a large scale field for the continuous dynamo process.\\

As $\mathbf{b}_{1}$ increases, the energy diffuses toward $\mathbf{b}_{3}$ through $-\mathbf{u}\cdot \nabla \mathbf{b}_{1}$ (Fig.~\ref{f3}). However, the direction of curl effect is opposite so the induced field heads for $-\hat{y}$. $\mathbf{b}_{3}(\hat{y})$ can be inferred to approach to $zero$ and regrow to be $\mathbf{b}'_{3}(-\hat{y})$. Also, the toroidal field $\mathbf{b}'_{tor}$ around $\mathbf{b}'_{3}$ is induced due to $\overline{\mathbf{j}}_{ind}$. The direction of $\mathbf{b}'_{tor}$ does not change. Subsequently, $\mathbf{b}'_{3}$ and $\mathbf{b}'_{tor}$ make the right handed magnetic helicity in small scale regime. This is supported by the changing sign of $H_M$ at minimum of $\mathbf{b}_3$ (see Fig.~\ref{f5b}). Simultaneously, $\mathbf{b}'_3$ can be suppressed by the interaction between $\mathbf{b}'_{tor}$ and $\mathbf{u}_i$. Also, the growth of $\mathbf{b}_{1}$ modifies the curvature radius of magnetic fields so that Lorentz force suppresses $\mathbf{u}_i$. All of these effects explain the conservation of magnetic helicity in HKFD.\\


Left panel in Fig.~\ref{f4a} shows the dynamo process discussed above. However, right panel shows a different possibility that $\mathbf{u}_{pol}\times \mathbf{b}_{0}$ can generate $\mathbf{j}_{ind,\,2}(-\hat{y})$ which is parallel to the $\mathbf{b}_{ind,\,2}\,(=\mathbf{b}_{ind,\,1})$. $\mathbf{j}_{ind,\,2}$ induces $\mathbf{b}_{tor,\,2}$, and $\mathbf{b}_{tor,\,2}$ forms the right handed magnetic helicity with $\mathbf{b}_{ind,\,2}$. These two dynamo processes seem to result in $zero$ net helicity. However, a careful look shows an essential difference between these two processes. The amplified $\mathbf{b}_{ind,\,1}$ due to the enhanced $\mathbf{b}_{0}$ yields stronger $\mathbf{j}_{ind,\,1}$, which amplifies $\mathbf{b}_{tor,\,1}$. This toroidal field fortifies $\mathbf{b}_0$ again, which produces the enhanced $\mathbf{b}_{ind,\,1}$. In contrast, $\mathbf{b}_{ind,\,2}$ and $\mathbf{b}_{tor,\,2}$ do not have such a mutual interaction, rather evolve passively. This essential difference decides the dominant magnetic helicity in the system.\\

The principle of helical magnetic forcing dynamo (HMFD) in Fig.~\ref{f4b} is similar to that of HKFD except some slight but essential differences. In the system forced by the right handed (positive) helical magnetic energy, $\mathbf{b}_{ind,\,3}(\hat{y})$ is generated from $\mathbf{u}_{0}\times \mathbf{b}_{tor}$. But, $\mathbf{u}_0$ can induce two different current densities: $\mathbf{j}_{ind,\,3}(\hat{y})$ from $\mathbf{u}_{0}\times\mathbf{b}_{pol}$ and $\mathbf{j}_{ind,\,4}(\hat{z})$ from $\mathbf{u}_{0} \times \mathbf{b}_{ind,\,4}\,(\mathbf{b}_{ind,\,3}=\mathbf{b}_{ind,\,4})$. $\mathbf{j}_{ind,\,3}(\hat{y})$ generates $\mathbf{b}_{tor,\,3}$ around $\mathbf{b}_{ind,\,3}$ leading to the right handed magnetic helicity. But, $\mathbf{j}_{ind,\,4}$ generates $\mathbf{b}_{tor,\,4}$ around $\mathbf{b}_{ind,\,4}$ to generate the left handed magnetic helicity. However, since $\mathbf{b}_{tor,\,4}$ is antiparallel to $\mathbf{b}_{tor}$, the left handed magnetic helicity cancels with the injected right handed one. This is why the magnetic helicity generated in HMFD has the same sign of a forcing magnetic helicity.\\

\subsection{Derivation of $\alpha$ coefficients}
Thus far, we have shown the physical mechanisms of LSD and SSD in the field structure. Being different from SSD, LSD requires an additional amplifying process of the $\mathbf{B}$ field, $\alpha$ effect. The model implies that the growth of $\overline{\mathbf{B}}$ field ($\mathbf{b}_1$) is related to the (helical) motion of $\mathbf{u}_i$, $\mathbf{b}_i$ in small scale and $\overline{\mathbf{B}}$ field itself. Therefore, if the characteristic length $l$ and time scale $\tau$ of turbulent eddies are smaller than those of $\overline{\mathbf{B}}$, EMF may be expanded like:
\begin{eqnarray}
\langle \mathbf{u}\times \mathbf{b}\rangle_i\sim
\alpha_{ij}\overline{B}_{j}+\beta_{ilm}\frac{\partial\, \overline{B}_{l}}{\partial\, x_m}+\gamma_{ilmn}\frac{\partial^2 \overline{B}_{l}}{\partial x_m\partial x_n}...
\label{alpha_derivation_1}
\end{eqnarray}
Moreover, if a smaller quantity out of $Re_M\equiv ul/\eta$ and $S\equiv u\tau/l$ is much smaller than `1' (min$(Re_M,\,S)\ll1$), triple correlation terms and $G$ $(\equiv (\mathbf{u}\times \mathbf{b}-\langle \mathbf{u}\times \mathbf{b}\rangle)$ in the magnetic induction equation in the small scale regime can be neglected. Then, $\xi$ can be calculated from $\mathbf{u}\times \int^{\tau}\partial \mathbf{b}/\partial t\,dt$ \citep{1978mfge.book.....M} or from $\int^{\tau}\partial \mathbf{u}/\partial t\,dt\times \mathbf{b}$ \citep{1983PhFl...26.2558K}. However, these anticommutative FOSAs are not generally valid besides the considerable $G$ in space. In MTA, the third order moment terms, neglected in FOSA, are replaced by $\xi/\tau$ without further calculation. It starts from the differentiation of a multi-variable function $\mathbf{\xi} (\mathbf{u},\, \mathbf{b})$.
\begin{eqnarray}
\frac{\partial }{\partial\, t}\nabla \times \langle \mathbf{u}\times \mathbf{b}\rangle=
\nabla \times \big\langle \frac{\partial \mathbf{u}}{\partial\, t}\times \mathbf{b}\big\rangle+
\nabla \times \big\langle \mathbf{u}\times \frac{\partial \mathbf{b}}{\partial\, t}\big\rangle.
\label{alpha_derivation_2}
\end{eqnarray}
After some analytical calculations, we can derive the simple forms of $\alpha=1/3\int^{\tau}(\langle \mathbf{j}\cdot \mathbf{b}\rangle-\langle \mathbf{u}\cdot \mathbf{\omega}\rangle)dt$ and $\beta=1/3\int^{\tau}\langle u^2\rangle dt$. Mathematically, the quenching effect of $\mathbf{j}\cdot \mathbf{b}$ in the $\alpha$ coefficient is from the definition of a vector product and differentiation of a multi variable function. Physically, it is caused by the interaction between the current density and magnetic field in different scales.\\

Additional differentiation of Eq.~(\ref{alpha_derivation_2}) produces the fourth order moments which can be decomposed into the combination of second order ones: $\langle X_1X_2X_3X_4\rangle\sim \sum_{ijlm}\langle X_iX_j\rangle\langle X_lX_m\rangle$ (QN, \cite{1967PhFl...10..859K}). With the assumption of isotropy without reflection symmetry, the second order moment can be replaced by
\begin{eqnarray}
\langle X_l(k)X_m(-k)\rangle&=&P_{lm}(k)E(k)+\frac{i}{2}\frac{k_n}{k^2}\epsilon_{lmn}H(k),
\label{second_order_moment_replacement}
\end{eqnarray}
where $P_{lm}(k)=\delta_{lm}-k_lk_m/k^2$, $\langle X^2/2\rangle=\int E(k) \,d\mathbf{k}$, and $\langle \mathbf{X}\cdot\nabla\times \mathbf{X}\rangle=\int H(k) \,d\mathbf{k}$. With some calculations, $\alpha$, $\beta$ coefficient similar to those of MTA can be derived \citep{1975JFM....68..769F}.\\

All these methods are essentially to solve a closure issue in the MHD equations approximately without the exact solution of anisotropy and energy cascade time $\tau$ affected by the magnetic field. Therefore, if there is a practical method to find the exact $\alpha$, $\beta$ coefficient from observation and simulation, it will be helpful to infer a better closing method leading to a more exact helical dynamo theory.\\

\subsection{Derivation of `$\alpha$' and `$\beta$' coefficients using semi-analytic method}
In principle, $\alpha$ and $\beta$ coefficients can be found from $\partial \overline{\mathbf{B}}/\partial t\sim \nabla \times \alpha \overline{\mathbf{B}} + (\eta+\beta) \nabla^2\overline{\mathbf{B}}$. However, this vector equation is not so useful to the practical calculation. Instead, the scalar equations for $\overline{E}_M$ and $\overline{H}_M$ are more useful. We can get them like \citep{2017MNRAS.472.1628P}
\begin{eqnarray}
\frac{\partial}{\partial t}\langle{\bf \overline{A}}\cdot {\bf \overline{B}}\rangle&=&
2\langle{\bf \overline{\xi}}\cdot {\bf \overline{B}}\rangle-2\eta \langle{\bf \overline{B}}\cdot{\bf \nabla} {\bf \times} {\bf \overline{B}}\rangle\nonumber\\
&=&2\alpha\langle {\bf \overline{B}}\cdot {\bf \overline{B}} \rangle-2(\beta+\eta)\langle {\bf \overline{B}}\cdot{\bf \nabla}{\bf \times}{\bf \overline{B}}\rangle\nonumber\\
\rightarrow \frac{\partial}{\partial t}\overline{H}_M&=&4\alpha \overline{E}_M-2(\beta+\eta)\overline{H}_M,\label{Hm1}\\\nonumber\\\nonumber\\
\frac{\partial }{\partial t}\frac{1}{2}\langle \overline{B}^2\rangle&=&\langle{\bf \overline{B}}\cdot{\bf \nabla}{\bf \times} \overline{{\bf\xi}}\rangle-\frac{c}{\sigma}\langle{\bf \overline{B}}\cdot {\bf \nabla}{\bf \times}{\bf \overline{J}}\rangle\nonumber\\
&=&\langle\alpha {\bf \overline{B}}\cdot {\bf \nabla} {\bf \times}{\bf \overline{B}}\rangle-\langle\beta {\bf \nabla} {\bf \times} {\bf \overline{B}}\cdot {\bf \nabla} {\bf \times}{\bf \overline{B}}\rangle-\frac{c}{\sigma}\langle{\bf \overline{J}}\cdot {\bf \nabla}{\bf \times}{\bf \overline{B}}\rangle.\nonumber\\
\rightarrow \frac{\partial }{\partial t}\overline{E}_M&=&\alpha \overline{H}_M-2\big(\beta+\eta\big)\overline{E}_M.
\label{Em1}
\end{eqnarray}
The solutions of coupled Eqs.~(\ref{Hm1}), (\ref{Em1}) are\footnote{$\overline{E}_M (\sim\langle \overline{B}^2 \rangle$) includes the helical and nonhelical components of $\overline{B}$. However, the nonhelical component of $\mathbf{B}$ is dropped when $ \overline{\mathbf{A}}\cdot \overline{\mathbf{B}}$ is averaged over the large scale.},
\begin{eqnarray}
2\overline{H}_M(t)&=&(\overline{H}_M(0)+2\overline{E}_M(0))e^{2\int^t_0(\alpha-\beta-\eta)d\tau}\nonumber\\
&+&(\overline{H}_M(0)-2\overline{E}_M(0))e^{-2\int^t_0(\alpha+\beta+\eta)d\tau},\label{Hm_final_solution}\\
4\overline{E}_M(t)&=&(\overline{H}_M(0)+2\overline{E}_M(0))e^{2\int^t_0(\alpha-\beta-\eta)d\tau}\nonumber\\
&-&(\overline{H}_M(0)-2\overline{E}_M(0))e^{-2\int^t_0(\alpha+\beta+\eta)d\tau}.\label{Em_final_solution}
\end{eqnarray}
Eqs.~(\ref{Hm_final_solution}), (\ref{Em_final_solution}) are consistent with the field structure (Fig.~\ref{f4a}, \ref{f4b}) and the simulation results (Fig.~\ref{f6a}, \ref{f6b}). If $\alpha <0$ (positive HKF or negative HMF), the second term in the right-hand side of Eqs.~(\ref{Hm_final_solution}), (\ref{Em_final_solution}) become dominant with increasing time. Since $2\overline{E}_M(0)\geq\overline{H}_M(0)$, $\overline{H}_M(t)$ becomes negative. In contrast, if $\alpha >0$ (negative HKF or positive HMF), $\overline{H}_M(t)$ becomes positive and converges to $2\overline{E}_M(t)$ eventually. But $\overline{E}_M(t)$ is positive in any case. When making these theoretical predictions, we referred to the simplest definitions of $\alpha$ and $\beta$ coefficients. However, these coefficients can be derived from
\begin{eqnarray}
2\int^t (\alpha-\beta-\eta)\,d\tau &=& ln\bigg(\frac{\overline{H}_M(t)+2\overline{E}_M(t)}{\overline{H}_M(0)+2\overline{E}_M(0)}\bigg),\\
2\int^t (-\alpha-\beta-\eta)d\,\tau &=& ln\bigg(\frac{\overline{H}_M(t)-2\overline{E}_M(t)}{\overline{H}_M(0)-2\overline{E}_M(0)}\bigg)
\label{EmHmSolution2}
\end{eqnarray}
With $C(t)\equiv \overline{H}_M(t)+2\overline{E}_M(t)$ and $D(t)\equiv \overline{H}_M(t)-2\overline{E}_M(t)$, we can further obtain the separate coefficients like
\begin{eqnarray}
\alpha(t)&=&\frac{1}{4}\bigg(\frac{1}{C(t)}\frac{\partial C(t)}{\partial t} - \frac{1}{D(t)}\frac{\partial D(t)}{\partial t}\bigg),\label{alphaSolution3}\\
\beta(t)&=&-\frac{1}{4}\bigg(\frac{1}{C(t)}\frac{\partial C(t)}{\partial t} + \frac{1}{D(t)}\frac{\partial D(t)}{\partial t}\bigg)-\eta.
\label{betaSolution3}
\end{eqnarray}
As the results show, if we know the temporal changes of large scale magnetic helicity $\overline{H}_M(t)$ and large scale magnetic energy $\overline{E}_M(t)$, $\alpha (t)$ and $\beta (t)$ coefficients can be found. Considering that other theoretical and numerical methods to find these coefficients require considerable analytic calculations and elaborate simulations, which are not yet accurate, these results give us quite a clear method to find $\alpha (t)$ and $\beta (t)$ coefficients. Additional differentiation over time leads to the integrand of each coefficients: $d\alpha(t)/dt=d\int^t \langle\cdot\rangle d\tau /dt$, $d\beta(t)/dt=d\int^t \langle\cdot\rangle d\tau /dt$. They can give us a chance to test the classical analytic results of FOSA or MTA $d\alpha(t)/dt\sim \langle \mathbf{j}\cdot \mathbf{b}\rangle - \langle \mathbf{u}\cdot \mathbf{\omega}\rangle$ or $d\beta(t)/dt\sim \langle u^2\rangle$. For a decaying MHD system, these classical results with Eq.~(\ref{Hm_final_solution}), (\ref{Em_final_solution}) reproduce the simulation results quite well \citep{2017MNRAS.472.1628P}. But still more numerical test is necessary for a forced system. There may be additional effects which are excluded in MTA and FOSA. Since the results Eq.~(\ref{alphaSolution3}), (\ref{betaSolution3}), which are mathematically derived with a statistical assumption and algebra, are exact as long as $\langle \mathbf{u}\times \mathbf{b}\rangle\sim \alpha \overline{\mathbf{B}}-\beta \nabla \times \overline{\mathbf{B}}$ is valid, we will be able to see which other terms play a role in the $\alpha$ effect.\\

\section{Summary}
Thus far, we have seen how the field structure model explains the amplification process of the magnetic field in the plasma. Magnetic field $\mathbf{b}_{nl}$ parallel to $\mathbf{u}$ is transferred through $\mathbf{b}\cdot\nabla \mathbf{u}$, and magnetic field $\mathbf{b}_{loc}$ parallel to $\mathbf{b}$ is transferred through $-\mathbf{u}\cdot\nabla \mathbf{b}$. The net magnetic field $\mathbf{b}_{net}$ from $\mathbf{b}_{nl}$ and $\mathbf{b}_{loc}$ is used as a seed magnetic field for the next dynamo step. As the field structure shows, growing $\mathbf{b}_{nl}$ parallel to $\mathbf{u}$ suppresses the dynamo process, whereas growing $\mathbf{b}_{loc}$ perpendicular to $\mathbf{u}$ boosts the dynamo action. This result explains the dependence of dynamo on the magnetic Prandtl number ($Pr_M\equiv\nu/\eta$). With less magnetic diffusion (decreasing $\eta$) and more kinetic dissipation (increasing $\nu$, or increasing $Pr_M$), the dynamo effect elevates. In contrast, the decreasing mechanical dissipation decreases the dynamo effect. These relations imply that the saturation of the magnetic field in an ideal system is related to the field structure between $\mathbf{u}$ and $\mathbf{b}$ (angle $\theta$) rather than the dissipation effect. We also explained the mechanism of the helical dynamo ($\alpha$ effect) using vector field analysis. HKFD and HMFD can generate both of the positive and negative magnetic helicity in principle. However, as we discussed, only opposite (same) sign of magnetic helicity is left in a forced HKFD (HMFD) system. Finally, we derived $\alpha$, $\beta$ coefficients from the large-scale magnetic energy and magnetic helicity. The exact coefficients are useful to understanding more accurate internal dynamo processes in a MHD system, which leads to a more general dynamo theory. At present, the field structure model with various physical conditions such as rotation, shear, or $\mathbf{B}_{ex}$ remains to be done. Before doing that, we will test the method with the simulation results and observation data.



\section*{Acknowledgements}
KWP appreciates the support from ERC Advanced Grant STARLIGHT: Formation of the First Stars (339177) and the support from bwForCluster for numerical simulation.




\bibliographystyle{mnras}
\bibliography{bibfile} 







\bsp	
\label{lastpage}
\end{document}